\documentclass[preprint]{aastex62}
\usepackage{amsmath,amstext}
\usepackage{graphicx}

\def\vec#1{\ensuremath{\mathbf{#1}}}

\received{2020 May 28}
\revised{2020 December 14}
\accepted{2021 January 2}

\shortauthors{Shi et al.}
\begin{document}

\title{The First 3D Coronal Loop Model Heated by MHD Waves against Radiative Losses}

\correspondingauthor{Mijie Shi}
\email{mijie.shi@kuleuven.be}

\author{Mijie Shi}
\affiliation{Shandong Key Laboratory of Optical Astronomy and Solar-Terrestrial Environment, Institute of Space Sciences, Shandong University, Weihai, Shandong, 264209, China}
\affiliation{Centre for mathematical Plasma Astrophysics, Department of Mathematics, KU Leuven, B-3001 Leuven, Belgium}

\author{Tom Van Doorsselaere}
\affiliation{Centre for mathematical Plasma Astrophysics, Department of Mathematics, KU Leuven, B-3001 Leuven, Belgium}

\author{Mingzhe Guo}
\affiliation{Shandong Key Laboratory of Optical Astronomy and Solar-Terrestrial Environment, Institute of Space Sciences, Shandong University, Weihai, Shandong, 264209, China}
\affiliation{Centre for mathematical Plasma Astrophysics, Department of Mathematics, KU Leuven, B-3001 Leuven, Belgium}

\author{Konstantinos Karampelas}
\affiliation{Centre for mathematical Plasma Astrophysics, Department of Mathematics, KU Leuven, B-3001 Leuven, Belgium}
\affiliation{Department of Mathematics, Physics and Electrical Engineering, Northumbria University, Newcastle Upon Tyne, NE1 8ST, UK}

\author{Bo Li}
\affiliation{Shandong Key Laboratory of Optical Astronomy and Solar-Terrestrial Environment, Institute of Space Sciences, Shandong University, Weihai, Shandong, 264209, China}

\author{Patrick Antolin}
\affiliation{Department of Mathematics, Physics and Electrical Engineering, Northumbria University, Newcastle Upon Tyne, NE1 8ST, UK}



\begin{abstract}
In the quest to solve the long-standing coronal heating problem, it has been suggested half a century ago that coronal loops could be heated by waves. Despite the accumulating observational evidence of the possible importance of coronal waves, still no 3D MHD simulations exist that show significant heating by MHD waves. Here we report on the first 3D coronal loop model heating the plasma against radiative cooling. The coronal loop is driven at the footpoint by transverse oscillations and subsequently the induced Kelvin-Helmholtz instability deforms the loop cross-section and generates small-scale structures. Wave energy is transfered to smaller scales where it is dissipated, overcoming the internal energy losses by radiation. These results open up a new avenue to address the coronal heating problem.
\end{abstract}




\section{Introduction} 
The mechanisms to heat the corona to its multi-million Kelvin temperature have been divided between DC heating mechanisms and AC heating mechanisms. The DC heating mechanisms are models generating currents and magnetic reconnection. Given that previous detection of coronal waves were sporadic events \citep{1999ApJ...520..880A,1999Sci...285..862N}, modelling efforts have mainly gone into 3D DC heating models \citep{2005ApJ...618.1020G,2014A&A...564A..12C,2016A&A...585A...4C}. However, the recent observational evidence for the ubiquitous presence of transverse coronal waves \citep{2007Sci...317.1192T,2011Natur.475..477M,2015A&A...583A.136A}, motivates a revisiting of the AC heating mechanisms.
 Wave heating has been successfully achieved in 1D \citep{2005ApJ...632L..49S} and
 reduced 3D MHD
 \citep{2011ApJ...736....3V,2014ApJ...787...87V} models.
 Recently, \cite{2018MNRAS.476.3328M} extended the 1D wave
 heating models to 3D, but the model lacked any transverse structuring
 and thus the waves heat a flux tube without a density-enhanced coronal loop.
 Recent 3D models with density-enhanced coronal loops have not been
 able to sustain the corona \citep{2018A&A...616A.125P,2019A&A...623A..37P}.
This is because the wave heating
is insufficient to compete against the strong radiative cooling of the coronal loops.
  Since the optically thin radiative cooling directly decreases the internal energy, it is the key quantity to overcome in coronal heating models. 
\par

The observed transverse waves that show little to no decay \citep{2012ApJ...759..144T,2012ApJ...751L..27W} are called decayless kink waves \citep{2013A&A...552A..57N}. Even though it is still unconfirmed whether these transverse waves have enough energy for heating the corona \citep{2007Sci...317.1192T,2011Natur.475..477M,2014ApJ...795...18V}, the observed energy flux of transverse oscillations is usually underestimated because some ``dark'' energy is hidden in the non-thermal broadening of spectral lines \citep{2012ApJ...761..138M,2012ApJ...746...31D,2019ApJ...881...95P}. The decayless waves are interpreted as being driven at the footpoints by photospheric motions, but the wave energy is transfered and dissipated in the corona, either through resonant absorption \citep{1988JGR....93.5423H,1992ApJ...384..348G}, mode coupling \citep{2012A&A...539A..37P}, phase mixing \citep{2017A&A...601A.107P}, or Kelvin-Helmholtz instability \citep{1984A&A...131..283B,2008ApJ...687L.115T,2015ApJ...809...72A}. The latter mechanism cascades the driving energy from large scales to small scales, where it is dissipated at low Reynolds numbers \citep{2017A&A...604A.130K,2019ApJ...870...55G,2019ApJ...876..100A,2019A&A...623A..53K}. The resulting loops have a fully turbulent cross-section \citep{2018A&A...610L...9K,2019FrP.....7...85A} and show decayless oscillations resembling observed events \citep{2019FrASS...6...38K}.
 
In this paper, we study the heating effects of transverse oscillations
on coronal loops via three dimensional MHD simulations.
Our work is different from the previous ones in the following three aspects:
1. We explicitly include the radiative cooling in simulations.
2. Our results show that the dissipation of transverse oscillations is promising to balance the radiative cooling of the coronal loops.
3. As far as we know, we propose a new method estimating numerical viscosity and resistivity in MHD simulations.
The numerical model is presented in Section \ref{numerical_model}.
Section \ref{results} provides the numerical results,
and Section \ref{summary_and_conclusion} is the discussion and conclusion.

\section{Numerical Model}
\label{numerical_model}
We model the coronal loop as a straight, density enhanced magnetic flux tube
	in the Cartesian coordinate.
Gravity is ignored.
The loop is along $z$ direction,
	with all physical parameters depending only on the transverse position $x$ and $y$.
Specifically, density is given by this relation,
\begin{eqnarray}
\rho = \rho_e + (\rho_i - \rho_e)*\zeta(x,y),\quad
\zeta(x,y) = \frac{1}{2}(1-{\rm{tanh}}(b(\sqrt{x^2+y^2}/R-1))),
\end{eqnarray}
where $[\rho_i,\rho_e] = [3\times10^8,10^8]m_p~{\rm cm^{-3}}$ denote the
	density in the interior and exterior of the loop,
	$R = 1~\rm{Mm}$ is the loop radius.
$b$ is a parameter controlling the width of the transition layer $l$,
	which we choose $b=20$, corresponding to $l\approx 0.3R$.
To study the heating of the loop we set temperature inside and outside 
	of the loop to be equal, $T_i = T_e = 1~\rm{MK}$,
	in order to avoid mixture of plasma with different temperatures.
The magnetic field is along $z$ direction,
	with the magnetic filed strength inside of the loop being 
	$B_i=30~\rm{Gauss}$.
Stability is ensured by maintaining the total pressure balance. 
The Alfv\'en speed inside of the loop reads 
	$v_{Ai}\approx 3778~\rm{km~s^{-1}}$.

Similar to previous works 
	\citep{2010ApJ...711..990P},
we trigger kink wave by implementing a 
	time-dependent transverse velocity driver at the footpoint. 
The velocity inside of the loop is 
\begin{eqnarray}
[v_x,v_y]=[v_0{\rm{cos}}(\frac{2\pi t}{P_k}),0],
\end{eqnarray}
and the velocity outside of the loop is 
\begin{eqnarray}
[v_x,v_y]=v_0{\rm{cos}}(\frac{2\pi t}{P_k})R^2[\frac{x^2-y^2}{(x^2+y^2)^2},\frac{2xy}{(x^2+y^2)^2}],
\end{eqnarray}
where $P_k$ is the period of the fundamental kink mode.
We choose the loop length $L = 200~\rm{Mm}$, 
	so that $P_k = 2L/c_k = 86.4~\rm{s}$, 
	with $c_k\approx 4627~\rm{km~s^{-1}}$ being the kink speed.
In our simulations we set 
$v_0 = 8~\rm{km/s}$, corresponding to the lower limits of chromospheric observations of the transverse waves \citep{2007Sci...318.1574D},
which is larger than previously used typical values at the photosphere \cite[e.g.][]{2017A&A...604A.130K}.
	
The simulations are performed using the ideal MHD module of the PLUTO code 
\citep{2007ApJS..170..228M}.
PLUTO solves the conservative form of the ideal MHD equations (i.e., continuity, momentum, induction, and energy equations).
We use the finite volume piecewise parabolic method (PPM) for spatial reconstruction,
and the second order Runge Kutta for time stepping.
The HLLD approximate Riemann solver is used for flux computation,
and a hyperbolic divergence cleaning method (GLM) is used to keep the magnetic field divergence-free.
The simulation domain is
$[-8,8]~\rm{Mm}$, $[-8,8]~\rm{Mm}$, and $[0,200]~\rm{Mm}$
in $x$, $y$, and $z$ directions.
To resolve the small-scale structures as far as possible in the transverse direction,
we use 800 uniform grid points in both the $x$ and $y$ directions,
and 100 uniform grids in the $z$ direction, corresponding to a spatial resolution in the $x$ or $y$ direction of $20~\rm{km}$.
For the side boundaries ($x$ and $y$), we use outflow boundary conditions
for all quantities.
The transverse velocities at the bottom boundary are given by the driver,
and at the top boundary are fixed to be zero.
At both the top and bottom boundary,
the $z$ component of velocity is antisymmetric,
while other quantities have zero-gradient boundary conditions.
In the simulations, we have ignored resistivity, viscosity, and thermal conduction. 
The energy dissipation in our simulation is entirely caused by numerical dissipation at the grid scales. 

\section{Results}
\label{results}
To study the heating effects of footpoint driven kink waves in the presence of radiative cooling, 
we perform a MHD simulation with the radiative cooling included (radiative simulation hereafter).
As the comparison case, an ideal MHD simulation (adiabatic simulation hereafter) similar to previous works
\citep{2017A&A...604A.130K,2019ApJ...870...55G} is also conducted.
The setup of the radiative simulation
is the same as the adiabatic simulation initially,
but with radiative cooling turned on when $t > 600~\rm{s}$.
The radiative cooling function $\Lambda(T)$ is tabulated from the Chianti 
database\footnote{https://www.chiantidatabase.org/}
\citep{2019ApJS..241...22D}.
At the same time when radiative cooling is turned on,  
in order to maintain the background corona,
we also turn on a temporally and spatially independent constant heating term across the whole domain 
$H_0 = n_e^2\Lambda_0$, where $n_e = 10^8~\rm cm^{-3}$, $\Lambda_0 = 2.6\times10^{-22}\rm~erg~cm^3~s^{-1}$.
$H_0$ is implemented to maintain the background corona via balancing its radiative loss.
By having both the radiative cooling and the background heating, we are in fact analyzing the heating effects after 
the KHI has fully developed
and only inside the density enhanced coronal loops.

\subsection{Loop Dynamics}
\begin{figure}
	\begin{center}
		\includegraphics[width=0.8\linewidth]{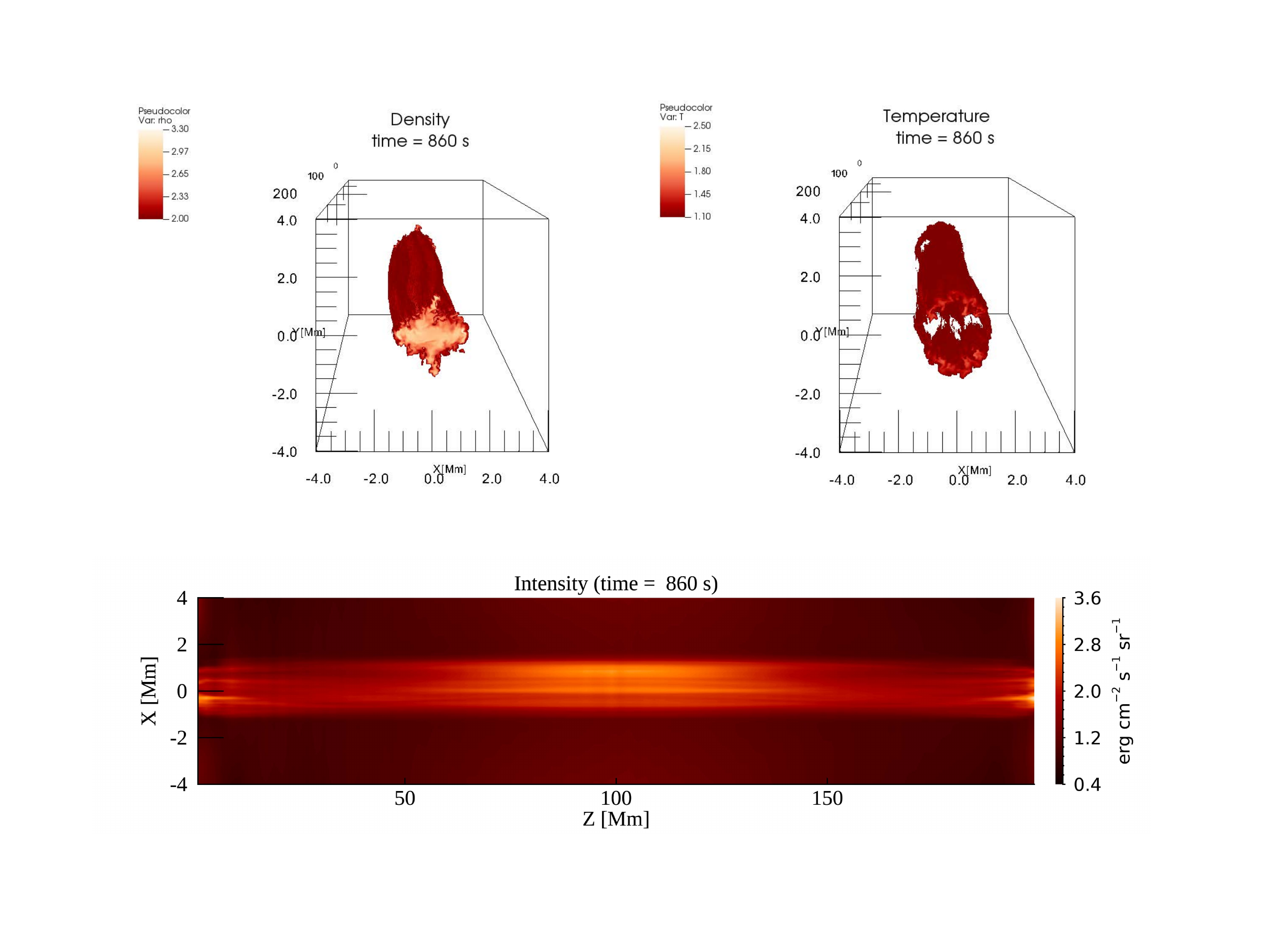}
		\caption{3D structure of the loop and the synthetic intensity at $t = 860~\rm{s}$ $(\sim10P_k)$ for radiative simulation.
			\textit{Top left:} 3D density structure of subvolume with $\rho \ge (\rho_e+\rho_i)/2$ (unit: $10^8~\rm{cm^{-3}}$). 
			\textit{Top right:} temperature structure with $T \ge 1.1 $ MK. 
			\textit{Bottom:} forward modeled (with the line of sight along $y$ axis) intensity of the Fe XII 193.5 \AA~line.}
		\label{f1}
	\end{center}
\end{figure}

For both cases, we run simulations till $t = 2000~\rm{s}$ $(\approx 23P_k)$.
Figure \ref{f1} displays the 3D density structure of the subvolume with $\rho \ge (\rho_e + \rho_i)/2$,
the temperature structure of the subvolume with $T \ge 1.1~\rm{MK}$,
and the forward modeled intensity of the Fe XII 193.5 \AA~line
by FoMo code\footnote{https://wiki.esat.kuleuven.be/FoMo}
\citep{2016FrASS...3....4V} for the radiative simulation
at $t = 860~\rm{s}$ $(\approx 10P_k)$.
From Figure \ref{f1} and the related movies we see the formation of a standing kink mode as the result of the footpoint driver.
The loop is distorted first at the boundary due to the velocity shear,
and then in the whole loop cross section as KHI develops.
Heating is initiated at the loop boundary and more profound at the loop boundary than the loop interior,
due to strong velocity shear at the boundary
and strong radiative loss at the loop interior.
Heating occurs first at the footpoint and only after at the loop apex,
making the intensity enhancement of the Fe XII 193.5 \AA~line (formation temperature $\sim$ 1.58 MK)
first at the footpoint and then at the loop apex.

\begin{figure}
	\begin{center}
		\includegraphics[width=\linewidth]{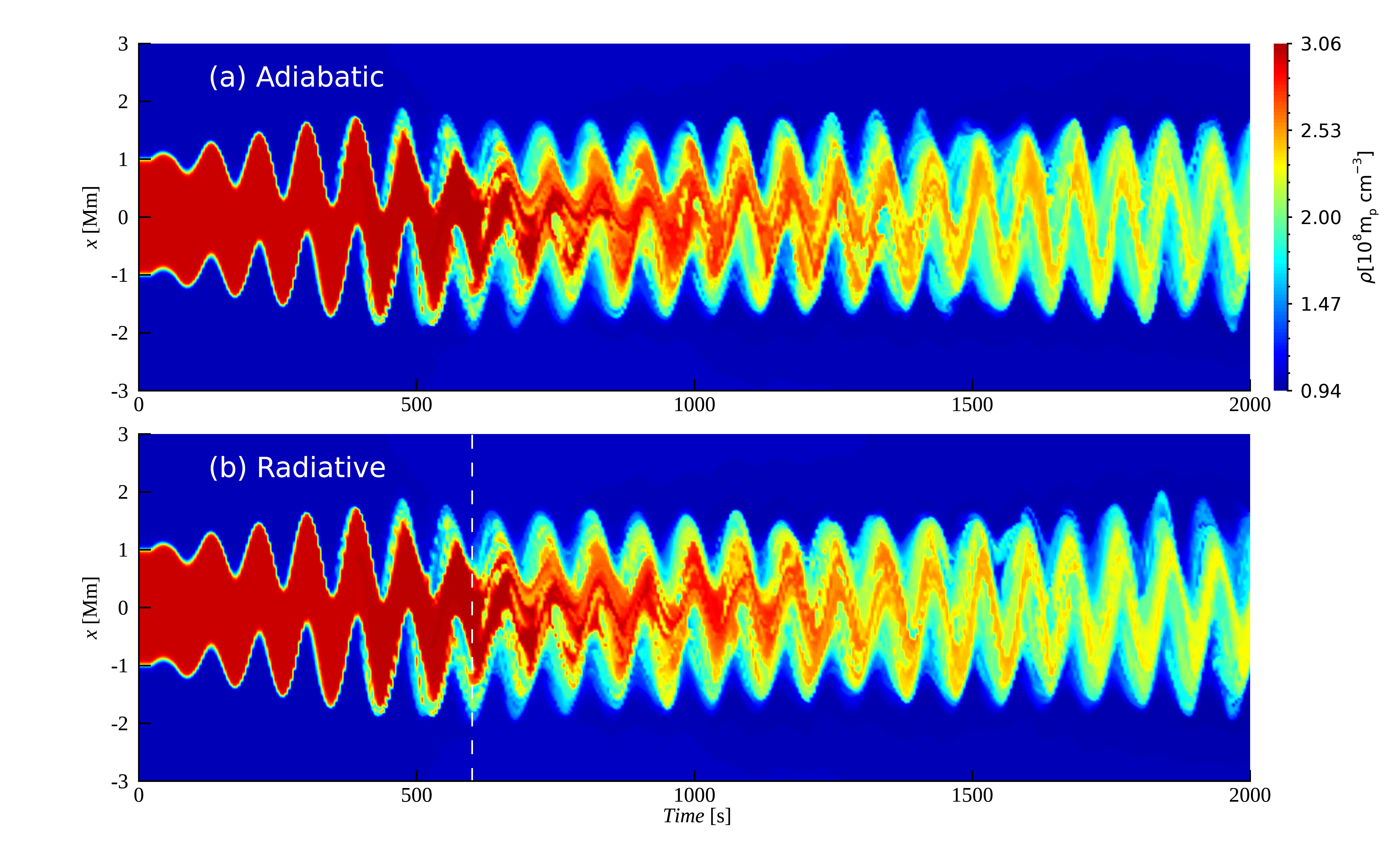}
		\caption{Time-distance maps of density at the
			mid-plane (a slice at $y=0$) of the loop apex for (a) adiabatic and (b) radiative simulations. Vertical dash line marks the time when radiative cooling is turned on.}
		\label{f1_revise}
	\end{center}
\end{figure}

Figure \ref{f1_revise} shows the time-distance maps of density at the
mid-plane (a slice at $y=0$) of the loop apex for two simulation cases.
We find that the introduction of the radiative cooling does not
have any obvious influence to the evolution of the wave amplitude.
The average displacement of the loop center is around 0.6 - 0.7 Mm in our simulations,
which is comparable to the larger amplitude limit 
of the observed decayless oscillations 
($\sim$ 0.5 Mm from \cite{2013A&A...552A..57N} and \cite{2015A&A...583A.136A}).

In order to examine the loop dynamics in more detail,
we plot in Figure \ref{f2} the density $\rho$, temperature $T$, 
$z$-component current $J_z$, and $z$-component vorticity $\omega_z$ 
at the loop apex $(z=0.5 L)$ and near the footpoint $(z=0.1 L)$ for both adiabatic and radiative simulations at 860 s ($\sim 10P_k$).
From Figure \ref{f2} and the related movie,
we find different evolutions at the apex and near the footpoint.
At the apex, the KHI is initiated at the loop boundary due to the strong velocity shear,
then develops further to the extent 
where the loop cross section is fully deformed.
Near the footpoint, where mixing is less strong,
the loop cross section is less deformed than at the apex.
As simulations go on, both $J_z$ and $\omega_z$ develop.
The high temperature regions coincide with the regions of 
strong current or viscosity.
$J_z$ is much stronger near the footpoint, 
while $\omega_z$ is more prominent around the apex,
indicating that numerical resistive heating is stronger near the footpoint,
while numerical viscous heating is more prominent around the apex
\citep{2007A&A...471..311V,2017A&A...604A.130K}.
Some high temperature peaks in the snapshot are also due to the adiabatic effects \citep{2017A&A...604A.130K}.
	Around the apex, mixing of different regions
	plays an important role redistributing and equilibrating the temperature,
	while near the footpoint, where mixing is less strong,
	the high temperature regions are more localized. 
The radiative cooling results in significantly lower temperatures in both the apex and near the footpoint. This is because the radiative cooling is proportional to the square of the density, and thus has most effect on the high density loop interior. However, despite the strong radiative cooling, also the radiative simulation maintains a temperature above the initial temperature of 1 MK. 
For the background corona, the gradual increase of the temperature around the apex is due to 
the ponderomotive force \citep{2004ApJ...610..523T},
which is also found in previous numerical simulations
\citep{2016A&A...595A..81M,2019A&A...623A..53K,2019ApJ...883...20G}.

\subsection{Volume-averaged Energy Density and Temperature}
In order to investigate the evolution of global energy and temperature,
we examine the volume-averaged energy density and temperature
in the subdomain of $|x|,|y| \le 4~\rm{Mm}$ 
for both simulations.
The variation of internal, kinetic, magnetic, and total energy densities with respect to their initial values ($E(0)$) are shown in the top panel of Figure \ref{f3}. 
In the adiabatic simulation,
the kinetic energy first increases as the loop is oscillating as a standing kink mode. 
When KHI has fully deformed the loop cross section by $t\approx 500~\rm{s}$ and small-scale structures are formed, 
the kinetic energy becomes saturated.
The numerical resistivity and viscosity increase the internal energy
and heat the plasma
via dissipating the magnetic and kinetic energy at the grid scales.
By inspecting the red solid line in the top panel of Figure \ref{f3},
we find that the evolution of internal energy has two stages.
When $t < 650~\rm{s}$, the increase rate of internal energy
is accelerating,
while after $t > 650~\rm{s}$, 
an almost constant increase rate is seen.
This difference is caused by the different dynamic behavior of the loop.
The first stage is during the growth stage of KHI,
with an accelerated energy transport from large to small scales.
During the second stage, 
when KHI has saturated and the loop cross section is fully deformed and small-scale structures are generated,
a constant energy transfer rate is found as indicated by the straight red solid line.

In the radiative simulation,
when radiative cooling is turned on at $t=600~\rm{s}$,
the kinetic energy is not obviously influenced,
while the magnetic and internal energies are changed compared to the adiabatic ones.
As the gas pressure of the loop is decreased by radiative cooling,
the inward Poynting flux at the side boundaries of the simulation increases the magnetic energy of the loop. 
Due to the cooling, 
the internal energy stops increasing and stabilises around a fixed value. Indeed, in this stage of our configuration, the driver energy input from the footpoint is approximately balancing the radiative energy losses. 
To better gauge the heating in the simulation, we examine the volume averaged temperature in the subdomain,
which is shown in the bottom panel of Figure \ref{f3}.
In the adiabatic case,
the temperature increases continuously due to persistent viscous and resistive heating at the small scales. It also follows the two stage process that we discussed before, with first a steady increase until the KHI saturates. In the latter stage, the temperature increase in the adiabatic case is nearly constant.
For the radiative simulation,
when radiative cooling is switched on,
the temperature remains almost constant in the second, radiatively cooling stage of the simulations.
This result demonstrates that the energy dissipation of small-scale structures can balance the 
radiative cooling and maintain a high temperature loop.
It also indicates that transverse oscillation dissipation is a promising mechanism for heating coronal loops.

\begin{figure}
	\begin{centering}
		\includegraphics[width=\linewidth]{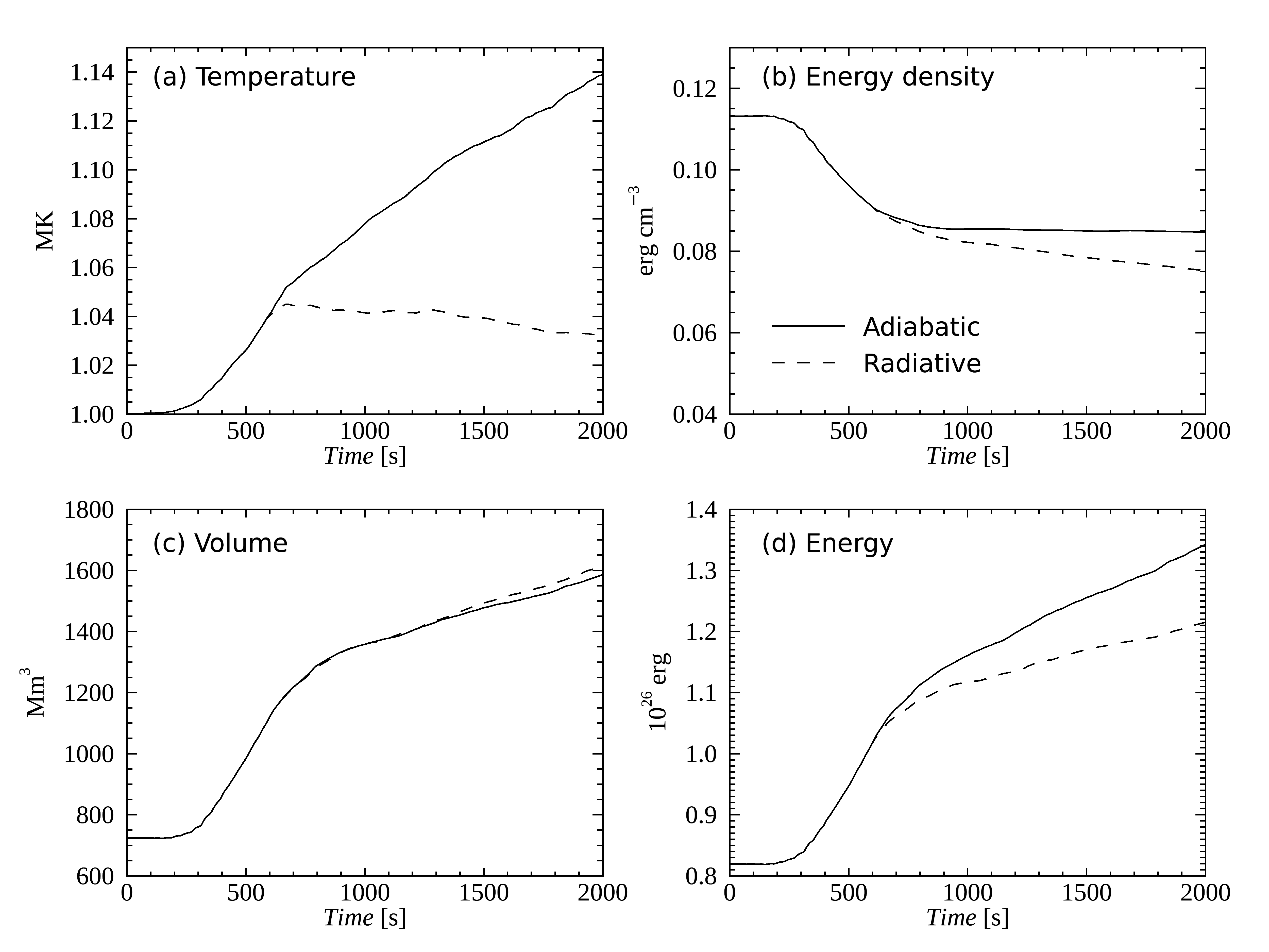}
		\caption{(a) Temperature, (b) Internal energy density, (c) volume, and (d) total internal energy of the domain with $\rho\ge1.1\rho_e$ for adiabatic (solid lines) and radiative (dashed lines) cases.}
		\label{f3_revise}
	\end{centering}
\end{figure}

To further address the heating effects,
it is important to show the internal energy evolution of the regions where radiative cooling is stronger than background heating.  
To do this, we select the regions with $\rho\ge1.1\rho_e$ and show in Figure \ref{f3_revise} the results of (a) volume averaged temperature, (b) volume averaged internal energy density, (c) total volume, and (d) total internal energy of this domain.
We find that the temperature within this domain shows similar trend as in Figure \ref{f3}, for both adiabatic and radiative cases.
This result shows that the heating effects are promising to balance the radiative cooling of the denser coronal loop.
However, the internal energy density decreases obviously, 
even in the adiabatic case,
which seems to contradict with the increase of temperature.
A similar situation was already discussed by 
\cite{2016A&A...595A..81M}
and \cite{2017A&A...604A.130K}.
The decrease of the internal energy density is not caused by the energy loss in this domain, but by the expansion of this domain,
as Figure \ref{f3_revise}(c) shows,
due to the development of KHI and the generated vortices.
The gradual increase of internal energy in Figure \ref{f3_revise}(d) is also predominantly caused by the expansion of the domain.

\subsection{Resistive and Viscous Heating Rates}
As mentioned in previous sections,
in our model energy dissipation and heating are caused by numerical dissipation.
Numerical resistivity or viscosity can be estimated by conducting a group of runs including varied physical resistivity or viscosity.
In this section, we estimate the numerical resistivity and viscosity using a different way by taking into account the energy equations.

\begin{figure}
	\begin{centering}
		\includegraphics[width=\linewidth]{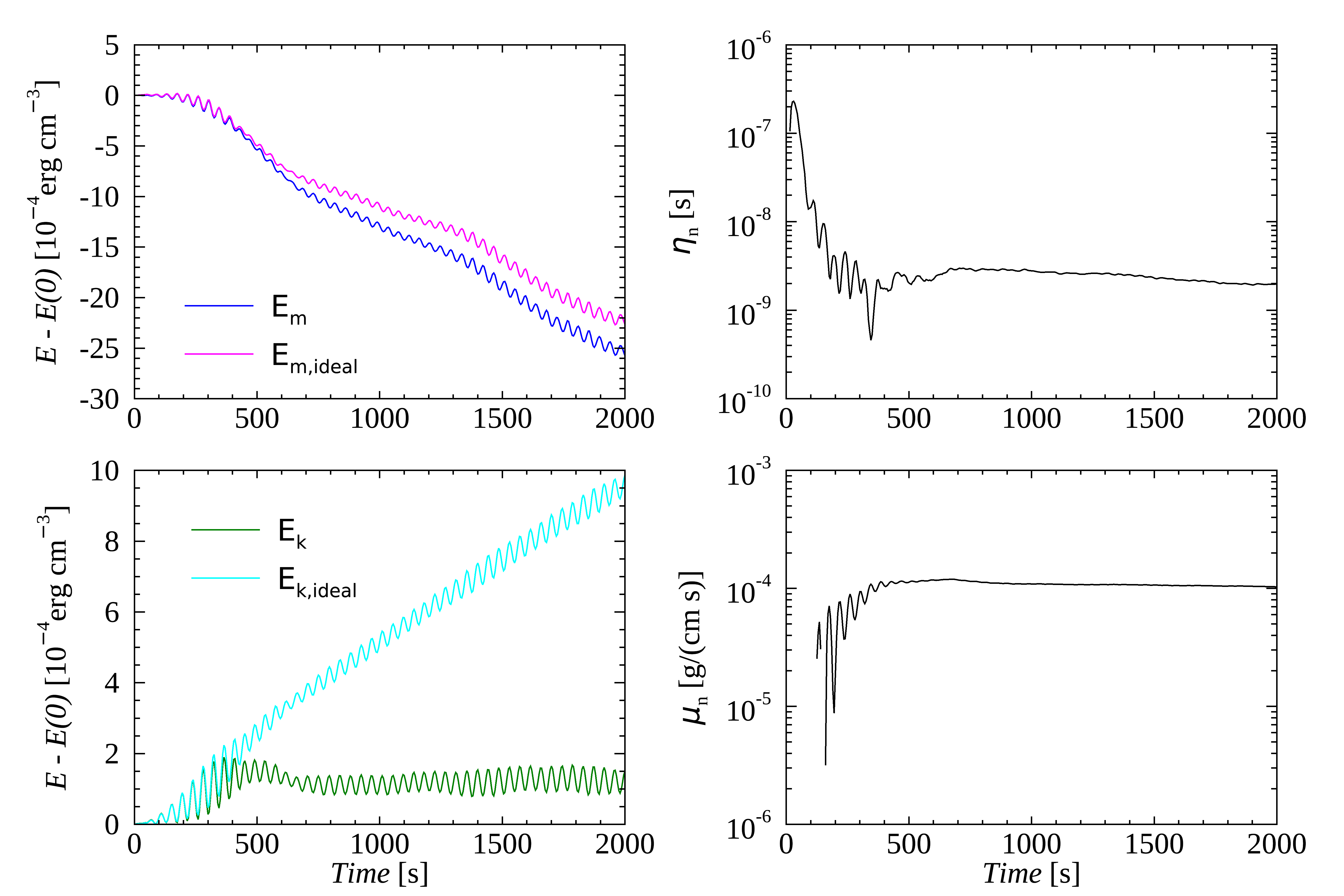}
		\caption{
			\textit{Top}: Evolution of magnetic energy density ($E_m$) and the magnetic energy density from ideal MHD ($E_{m,\rm{ideal}}$). Estimated numerical resistivity $\eta_n$.
			\textit{Bottom}: Evolution of kinetic energy density ($E_k$) and the kinetic energy density from ideal MHD ($E_{k,\rm{ideal}}$). Estimated numerical dynamic viscosity $\mu_n$. The results are for the adiabatic case in the fixed volume of $|x|,|y| \le 4~\rm{Mm}$.
		}
		\label{f_resis_vis}
	\end{centering}
\end{figure}

The change of volume-averaged magnetic energy density $B^2/8\pi$ can be expressed as \citep[e.g.,][]{2009LNP...780...43S},
\begin{eqnarray}
\begin{aligned}
\frac{1}{V_0}\int\frac{B^2}{8\pi}dV  &=
-\frac{1}{V_0}\iint\frac{c}{4\pi}(\vec{E}\times\vec{B})\cdot d\vec{S}dt - 
\frac{1}{V_0}\iint\vec{J}\cdot\vec{E}dVdt \\
 & = -\frac{1}{V_0}\iint\frac{c}{4\pi}[(\vec{E}_0+\eta_n\vec{J})\times\vec{B}]\cdot d\vec{S}dt - 
\frac{1}{V_0}\iint(\vec{J}\cdot\vec{E}_0+\eta_nJ^2)dVdt,
\label{eq4}
\end{aligned}
\end{eqnarray}
where we let $\vec{E} = \vec{E}_0 + \eta_n\vec{J}$.
$\vec{E}_0 = -(\vec{v}\times\vec{B})/c$ is the convective electric field and $\eta_n$ is the numerical resistivity.
For simplicity, we define the volume-averaged magnetic energy density as
\begin{eqnarray}
E_m = \frac{1}{V_0}\int\frac{B^2}{8\pi}dV.
\end{eqnarray}
Equation \ref{eq4} shows that the change of $E_m$ is caused by the terms from ideal MHD and the terms containing the numerical resistivity $\eta_n$.
We define the change of the magnetic energy density caused by ideal MHD as
\begin{eqnarray}
E_{m,\mathrm{ideal}}=-\frac{1}{V_0}\iint\frac{c}{4\pi}(\vec{E}_0\times\vec{B})\cdot d\vec{S}dt - 
\frac{1}{V_0}\iint\vec{J}\cdot\vec{E}_0dVdt.
\end{eqnarray}
Assuming $\eta_n$ is a constant,
$\eta_n$ can thus be derived by comparing $E_m$ and its ideal MHD counterpart $E_{m,\rm{ideal}}$.
The top panels of Figure \ref{f_resis_vis} show the results for the adiabatic case.
We find that initially $E_m$ matches with $E_{m,\rm{ideal}}$, but later on obvious difference is seen between the two curves.
This is because numerical resistivity will play a role dissipating magnetic energy when small-scale structures are generated.
We see that $\eta_n$ remains almost constant after $t\ge 500$ s.
The averaged numerical resistivity in the range of $500~\rm{s}\le t\le 2000~\rm{s}$ is $\eta_n =2.5\times10^{-9}~\rm{s}$.
The corresponding magnetic Reynolds number is 
$R_m = 4\pi v_cl_c/(c^2\eta_n) = 2.1\times10^5$,
taking the characteristic velocity as the initial internal Alfv\'en speed $v_c = v_{Ai} = 3778~\rm{km/s}$,
and the characteristic length as the loop radius $l_c = R = 1~\rm{Mm}$.

The change of volume-averaged kinetic energy density $\rho v^2/2$ is,
\begin{eqnarray}
\begin{aligned}
\frac{1}{V_0}\int\frac{1}{2}\rho v^2dV  &=
-\frac{1}{V_0}\iint[(\frac{1}{2}\rho v^2\bar{\bar{I}} + \bar{\bar{P}})\cdot\vec{v}]\cdot d\vec{S}dt + 
\frac{1}{V_0}\iint(p\nabla\cdot\vec{v} + \vec{J}\cdot\vec{E}_0 - \bar{\bar{\Pi}}:\nabla\vec{v})dVdt \\
\label{eq5}
\end{aligned}
\end{eqnarray}
where $\bar{\bar{P}} = p\bar{\bar{I}} - \bar{\bar{\Pi}}$ is the total stress tensor, and $\bar{\bar{\Pi}}$ is the viscous stress tensor.
For simplicity, we adopt the expression of viscous stress tensor from fluid dynamics \citep[e.g.,][]{2000ifd..book.....B}, $\Pi_{ij} = \mu_n(\frac{\partial u_i}{x_j} + \frac{\partial u_j}{x_i} - \frac{2}{3}\delta_{ij}\nabla\cdot\vec{v})$.
Here we use $\mu_n$ to denote the numerical dynamic viscosity.
Using a similar method above estimating $\eta_n$, 
and assuming $\mu_n$ is also a constant,
$\mu_n$ can be derived by comparing the kinetic energy density $E_k$ (left side of Eqution \ref{eq5}) and its ideal MHD counterpart $E_{k,\rm{ideal}}$ (right side of Equation \ref{eq5} without the $\bar{\bar{\Pi}}$ terms).
The bottom panels of Figure \ref{f_resis_vis} show the results.
We find that the kinetic energy density saturates at around $t = 500$ s, while its ideal MHD counterpart keeps increasing. The discrepancy between the two curves is caused by numerical viscosity.
The averaged numerical viscosity in the range of $500~\rm{s}\le t \le 2000~\rm{s}$ is $\mu_n = 1.1\times10^{-4}~\rm{g/(cm~s)}$.
The corresponding Reynolds number is
$R = l_cv_c/(\mu_n/\rho_0) = 1.7\times10^5$, taking $\rho_0 = 3\times10^8m_p~\rm{cm^{-3}}$ as the density of the loop interior at the initial state.

\begin{figure}
	\begin{centering}
		\includegraphics[width=\linewidth]{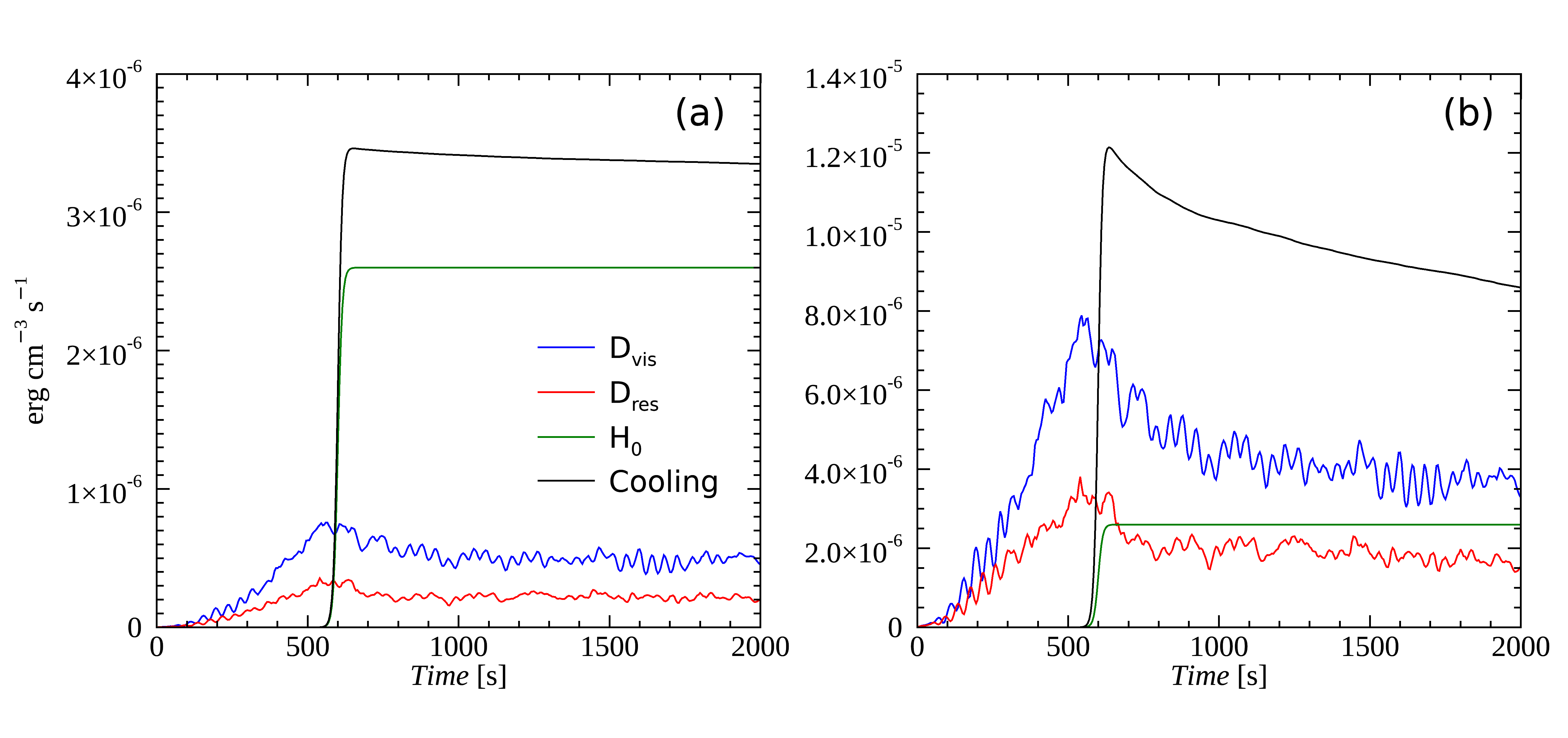}
		\caption{
			Viscous dissipation rate $D_{vis} = \bar{\bar{\Pi}}:\nabla\vec{v}$, resistive dissipation rate $D_{res} = \eta_nJ^2$, background heating rate $H_0$, and radiative cooling rate averaged in the domain of (a) $|x|,|y| \le 4~\rm{Mm}$, and (b) $\rho\ge 1.1\rho_e$ for the radiative case. Note: as stated in the text, both background heating and radiative cooling are smoothly turned on at 600 s.}
		\label{f_heating}
	\end{centering}
\end{figure}

Using the numerical resistivity $\eta_n$ and numerical dynamic viscosity $\mu_n$ from Figure \ref{f_resis_vis}, we can quantify the dissipation or heating rates caused by numerical resistivity $D_{res} = \eta_nJ^2$ and numerical viscosity $D_{vis} = \bar{\bar{\Pi}}:\nabla\vec{v}$.
Figure \ref{f_heating}(a) shows $D_{res}$, $D_{vis}$,
together with the background heating rate $H_0$, and the radiative cooling rate, averaged over the domain of $|x|,|y| \le 4~\rm{Mm}$ for the radiative case.
Both the background heating and the radiative cooling are smoothly turned on at $t = 600$ s.
From Figure \ref{f_heating}(a) we see that the cooling rate and background heating rate $H_0$ are several times larger than both $D_{vis}$ and $D_{res}$. This is mainly because both resistive and viscous dissipations are localized around the loop while Figure \ref{f_heating}(a) is averaged over a domain much larger than the loop size. 
$H_0$ is constant, as we mentioned above.
The radiative cooling rate is roughly balanced by $D_{vis}$, $D_{res}$, and $H_0$.
The radiative cooling rate decreases only slightly, 
meaning that the development of small-scale structures does not obviously influence the radiative loss of this domain.
Figure \ref{f_heating}(b) shows the same quantities but averaged over the domain of $\rho\ge1.1\rho_e$. This domain is not fixed, rather it expands as small-scale structures are generated (see Figure \ref{f3_revise}(c)).
The volume expansion obviously decreases the volume-averaged radiative cooling rate.
The radiative cooling rate is also roughly balanced by 
$D_{vis}$, $D_{res}$, and $H_0$.
Even though $H_0$ contributes around 30\% of the total heating rate in the time range of $700~\rm{s}\le t \le 2000~\rm{s}$,
the dissipations from resistivity and viscosity play the major role in the dynamics and deserve further future studies.

\section{Discussion and Conclusion}
\label{summary_and_conclusion}
	It is necessary to have some discussion about the thermal conduction, which is not included in our simulations.
	Thermal conduction can also cause energy loss of a loop. 
	In our initial setup, the time scales of a loop cooled down by thermal conduction ($\tau_c$) and by radiative cooling ($\tau_r$) can be roughly estimated by
	\citep[e.g.,][]{1994ApJ...422..381C}
	\begin{eqnarray}
	\tau_c = 4\times10^{-10} \frac{nL_{0}^2}{T^{5/2}} = 1.2\times10^4~\rm{s}
	\end{eqnarray}
	
	\begin{eqnarray}
	\tau_r = \frac{2nk_BT/(\gamma-1)}{n^2\Lambda(T)} = 5300~\rm{s}
	\end{eqnarray}
	where $n=3\times10^8~\rm{cm^{-3}}$ is the loop density, $L_0 = 100~\rm{Mm}$ is the loop half length, $T = 1$ MK, and $\Lambda(T) = 2.6\times10^{-22}~\rm{erg~cm^3~s^{-1}}$ is the radiative cooling rate at 1 MK.
	We have $\tau_c > \tau_r$ in our setup, indicating radiative cooling is more important than thermal conduction, though further analysis is required.
	
\begin{figure}
	\begin{centering}
		\includegraphics[width=\linewidth]{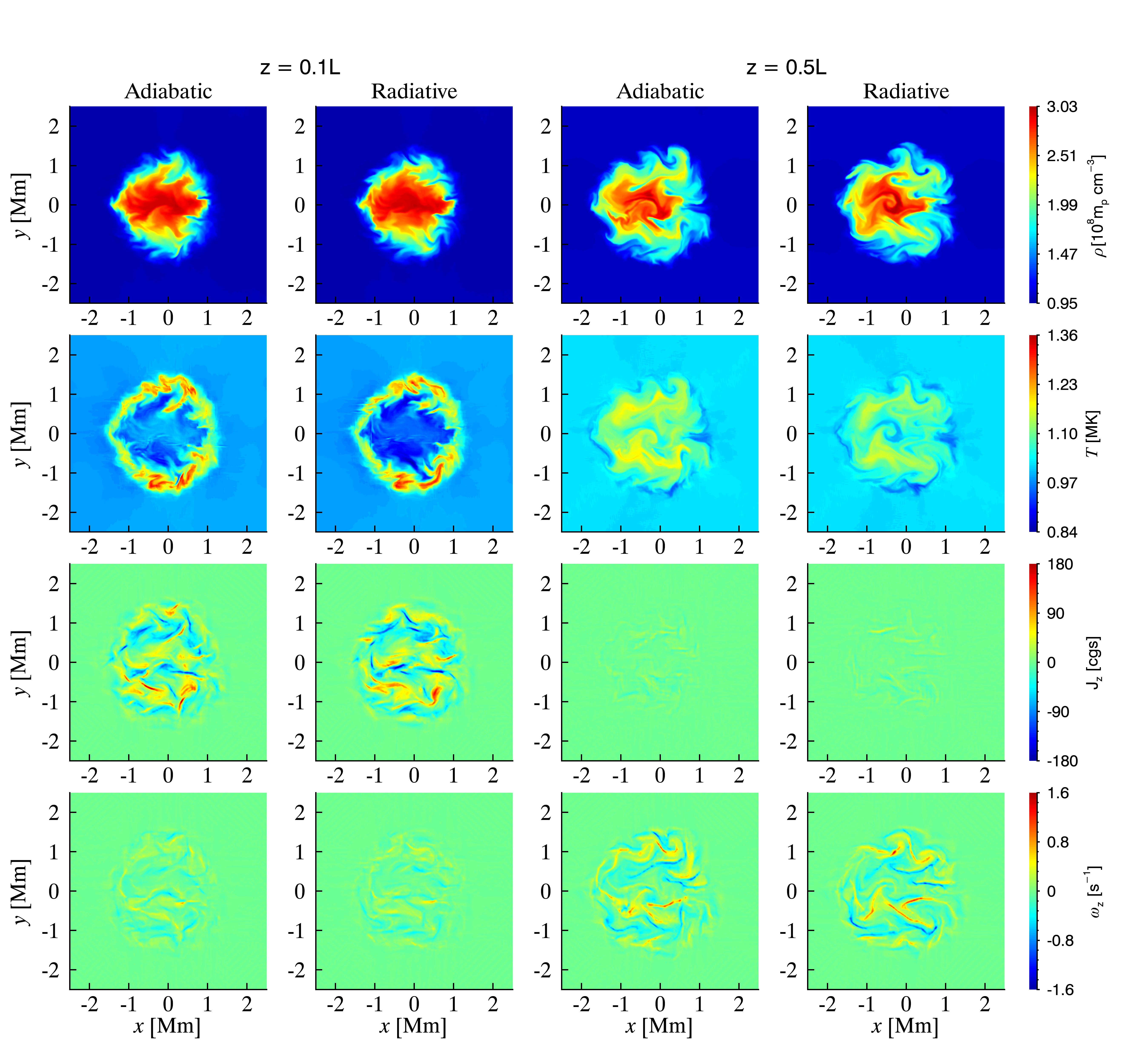}
		\caption{From top to bottom: plane cut of density, temperature, $z$-component of current and vorticity at $z = 0.1L$ and $z = 0.5L$ at $t=860$ s for adiabatic and radiative simulations.}
		\label{f2}
	\end{centering}
\end{figure}

\begin{figure}
	\begin{centering}
		\includegraphics[width=0.8\linewidth]{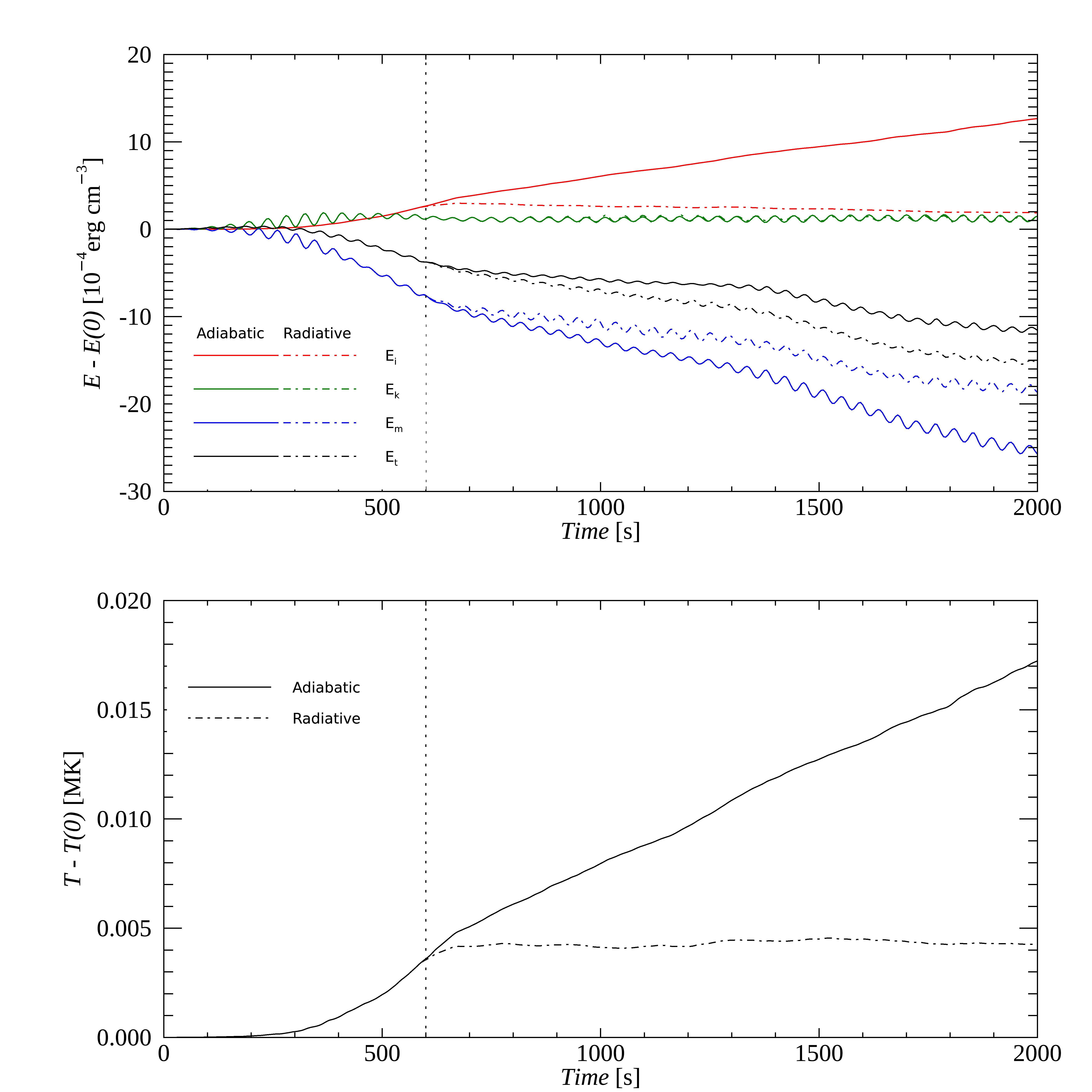}
		\caption{\textit{Top:} Volume averaged energy density variation of the internal (red),
			kinetic (green), magnetic (blue), and total (black) energy
			for the adiabatic (solid lines) and radiative (dash-dot lines) simulations.
			\textit{Bottom:} Volume averaged temperature variation of adiabatic (solid line)
			and radiative (dash-dot line) simulations.
			Vertical dot lines mark the time when radiative cooling is switched on. All quantities are averaged in the domain of $|x|,|y| \le 4~\rm{Mm}$.}
		\label{f3}
	\end{centering}
\end{figure}

\begin{figure}
	\begin{centering}
		\includegraphics[width=0.8\linewidth]{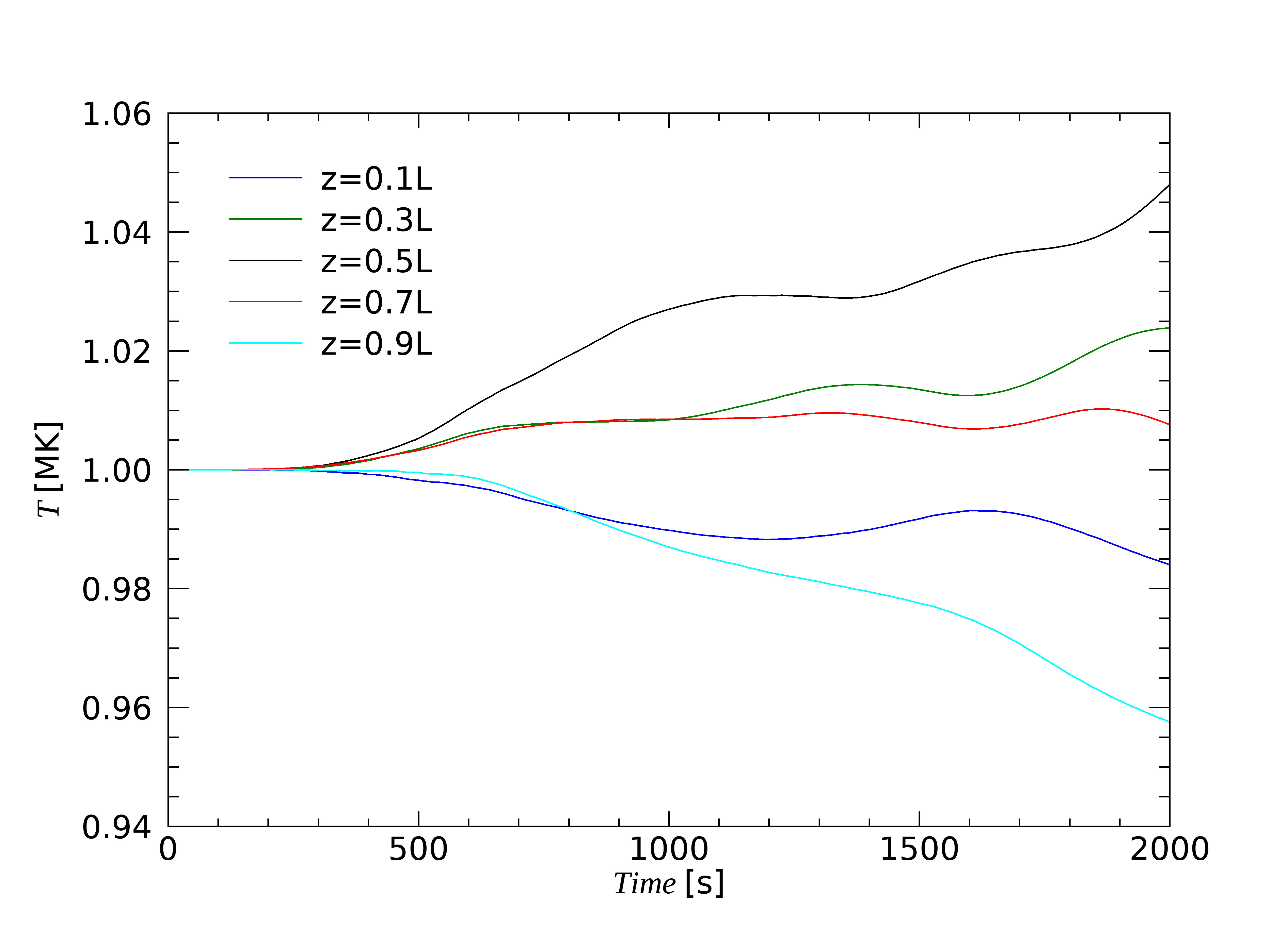}
		\caption{Temperature averaged over volumes centered at different positions of $z$ for the radiative case.}
		\label{f9}
	\end{centering}
\end{figure}

	Thermal conduction can transport heat along the magnetic field, which may change the temperature profiles we obtained in this work. Figure \ref{f3} shows that heating can globally balance cooling, while Figure \ref{f2} shows different temperature profiles around the footpoint and the apex.
	In Figure \ref{f9} we show the temperatures averaged over five subdomains with the same volume but centered at different $z$ positions for the radiative case.
	We find that heating is profound around the apex, while around the footpoint the temperature decreases because of cooling.
	Around $z=0.3L$ and $z=0.7L$, heating and cooling are almost balanced with each other.
	If thermal conduction is included, 
	the temperature profile along $z$ direction could be smoothed.
	In the adiabatic stage of $t < 600$ s, 
	we obtain a similar temperature profile as \cite{2019ApJ...883...20G} in their monolithic case. 
	The small reduction of the temperature near the footpoint
	is probably due to the ponderomotive force that can transport mass from the footpoint to the apex and thus influence the temperature profiles \citep{2004ApJ...610..523T}.

Thus, we have proven with our simulations that AC heating mechanisms can be used in 3D MHD models of coronal loops to heat the coronal plasma compensating for the massive radiative losses. This was done by comparing an adiabatic simulation with a simulation in which radiative cooling was switched on. In the latter simulation, the temperature was kept at coronal values, because of the balance between the dissipation of the input driver flux, the background heating, and the radiative cooling.\par
This model constitutes a proof-of-principle for the AC heating mechanisms which have been proposed half a century ago. Our model opens the door for further exploration of these heating mechanisms. This includes models incorporating a chromosphere and transition region who will induce field-aligned mass flows in response to the turbulent heating and thermal conduction. Moreover, we can look forward to models encompassing a whole active region heated with MHD waves.
More importantly, the models that will build upon our model can be contrasted to earlier DC heating models for the solar corona. This comparison could in principle indicate which observable effects that can distinguish the two heating classes in observations. 

Before closing, some remarks are necessary for future studies.
Firstly, our current simulations have energy outflows near the boundary, which could be improved in the future, 
and secondly, a parametric study where the wave period is changed spanning and bridging the DC AC transition would be helpful in determining which heating mechanism is more efficient at least in this model.

\acknowledgments
We thank the referees for the comments which have greatly improved our manuscript.
This work is supported by the European Research Council (ERC)
under the European Union’s Horizon 2020 research and innovation
program (grant agreement No. 724326),
the National Natural Science Foundation of China
(41904150, 11761141002, 41674172).
KK is supported by UK Science and Technology Facilities Council (STFC) grant ST/T000384/1 and by a FWO (Fonds voor Wetenschappelijk Onderzoek Vlaanderen) postdoctoral fellowship (1273221N). KK was also supported by a postdoctoral mandate from KU Leuven Internal Funds (PDM/2019).
PA acknowledges funding from his STFC Ernest Rutherford Fellowship (No. ST/R004285/2).
MS is support by the international postdoctoral exchange fellowship from China Postdoctoral Council.
CHIANTI is a collaborative project involving the University of Cambridge (UK), the NASA Goddard Space Flight Center (USA), the George Mason University(GMU, USA) and the University of Michigan (USA).
The computational resources and services used in this work were provided by the VSC (Flemish Supercomputer Center), funded by the Research Foundation Flanders (FWO) and the Flemish Government – department EWI.

\bibliographystyle{apj}
\bibliography{library}



\clearpage

\clearpage

\clearpage

\clearpage

\clearpage

\clearpage

\clearpage

\clearpage

%

\end{document}